\documentclass[fleqn,usenatbib]{mnras}

\usepackage[T1]{fontenc}
\usepackage{ae,aecompl}

\usepackage{amsmath}
\usepackage{amstext}
\usepackage{array}
\usepackage{graphicx}
\graphicspath{ {images/} }
\usepackage{natbib}
\usepackage{caption}
\usepackage{subcaption}
\usepackage{ulem}
\usepackage{amssymb}	

\author[J. N. Larson et al.]{
Jennifer N. Larson,$^{1}$\thanks{E-mail: jlarson27@Knights.ucf.edu (JNL)}
and G. Sarid,$^{2}$
\\
$^{1}$Dept. of Physics, University of Central Florida, Orlando, FL 32816, USA\\
$^{2}$SETI Institute, Mountain View, CA 94043, USA\\
}

\title[N-Body Debris Cloud Model]{An {\it N-Body} Approach to Modeling Debris and Ejecta Off Small Bodies: Implementation and Application}

\date{Accepted 2021 February 1. Received 2021 January 29; in original form 2019 November 11.}

\pubyear{2021}

\begin{document}
\label{firstpage}
\pagerange{\pageref{firstpage}--\pageref{lastpage}}
\maketitle

\begin{abstract}
We introduce here our new approach to modeling particle cloud evolution off surface of small bodies (asteroids and comets), following the evolution of ejected particles requires dealing with various time and spatial scales, in an efficient, accurate and modular way. In order to improve computational efficiency and accuracy of such calculations, we created an N-body modeling package as an extension to the increasingly popular orbital dynamics N-body integrator {\it Rebound}. Our code is currently a stand-alone variant of the {\it Rebound} code and is aimed at advancing a comprehensive understanding of individual particle trajectories, external forcing, and interactions, at the scale which is otherwise overlooked by other modeling approaches. The package we developed -- Rebound Ejecta Dynamics (RED) -- is a Python-based implementation with no additional dependencies. It incorporates several major mechanisms that affect the evolution of particles in low-gravity environments and enables a more straightforward simulation of combined effects. We include variable size and velocity distributions, solar radiation pressure, ellipsoidal gravitational potential, binary or triple asteroid systems, and particle-particle interactions. In this paper, we present a sample of the RED package capabilities. These are applied to a small asteroid binary system (characterized following the Didymos/Dimorphos system, which is the target for NASA's Double Asteroid Redirection Test mission).
\end{abstract}




\section{Introduction} \label{sec:intro}
Ejecta off small bodies is a fundamental consequence of processes that activate and modify surfaces of those small bodies. Recent reviews of small body impact and ejecta processes include those by \cite{Scheeres2002}, \cite{Housen2011}, and \cite{Jutzi2015}. With OSIRIS-Rex's discovery of frequent particle ejection events from asteroid Bennu \citep{Chesley2020}, and with active impact experiments such as those by Deep Impact \citep{AHearn2005,Richardson2007} and DART \citep{cheng2016}, understanding the role of ejecta in small-body evolution has taken on added significant currency. It can be driven by internal processes, like sublimation, rotational fission or thermal fatigue, but is more commonly associated with external forcing applied during impact events. Studies of post-impact debris clouds benefit investigations into ring formation, particle size distributions, cometary jets, and ejecta distributions. Many numerical models of impact ejecta are based on scaling relations by \cite{Housen1983}, \cite{Holsapple1993}, \cite{Richardson2007}, and \cite{Housen2011}. Scaling relations are a result of numerical models combined with experimental results. These relations are then used in other models, in order to include this process without having to calculate it explicitly. 

Hydrodynamic models estimate the flow of material in a cloud as a gradient. An example of this method can be found in the Maxwell Z Model \citep{Maxwell1977}, which estimates the flow of material out of a crater through streamlines. Computational models that implement a hydrodynamic approach, including iSALE-2D \citep{Amsden1980,Wunnemann2006,Collins2004,Ivanov1997,Melosh1992}, iSALE-3D \citep{Elbeshausen2009,Elbeshausen2011}, and FLAG \citep{burton1991,caramana1998}, focus primarily on the shock physics and fluid dynamics of the cratering processes. \cite{Richardson2007} updates the analytic models by \cite{Housen1983} to improve on interactions near the crater rim. \cite{Richardson2011} then applies these updated scaling laws to various 2D and 3D impact scenarios to model the flow of material from a crater to escape or back to the surface. The mathematical model by \cite{Richardson2007} sets up an analytical foundation upon which a dynamical framework using tracer particles to follow material flow is built. By using tracer particles to follow material flow using the streamline method, \cite{Richardson2011} analyzes how material is transported out of the impact crater and is deposited on the surface, creating a result similar to what is seen in N-body simulations.

These models are beneficial for determining the flow of material and gradient changes (e.g., temperature profiles, pressure, shock, etc.); however, hydrodynamic models lack the particle nature that can trace individual interactions within a cloud. While it is possible to include tracer particles that simulate a particle distribution, a full N-body integration can calculate the trajectories of individual particles outside of a grid or gradient. This is advantageous since it would allow for a detailed investigation of how particles of different size and different properties behave after ejection, thus providing us more insight into the specific processes at play. In addition, while hydrodynamic models typically simulate the impact cratering event itself, we are primarily interested in the post-impact debris dynamics. This added flexibility means that N-body code can take as inputs any conditions immediately after the event that produced the ejecta in the first place. 

Here we choose to use a new N-body code called {\it Rebound} \citep{ReinLiu2012} to examine the evolution of ejecta clouds off small bodies. We call our implementation of {\it Rebound} the ``Rebound Ejecta Dynamics package," or ``RED." The {\it Rebound} python module is an N-body code with both symplectic and non-symplectic integrators as well as a newly developed hybrid integrator (similar to the hybrid integrator used by MERCURY \citep{Chambers1997}, an N-body integrator that excels at large scale orbital evolution calculations) that switches between symplectic and non-symplectic integrators based on the a pre-defined condition of the system. The {\it Rebound} integrators have been validated with comparisons to other N-body integrators \citep{Silburt2016} and the demonstration of a multitude of scenarios in the {\it Rebound} simulation archive \citep{Rein2017}. We choose the Python {\it Rebound} module over other N-body integrators for this study due to its robustness and ability to carry out higher level computations at a lower performance cost than other N-body integrators currently being used. The two most commonly used N-body integrators for ejecta dynamics are SyMBA \citep{Duncan1998} and pkdgrav \citep{Stadel2001}, so we are building on earlier work that has successfully used N-body code in this physical context. Another popular N-body package capable of integrating complex systems is SyMBA \citep{Duncan1998}. While SyMBA does not have a specific hybrid integrator, it builds shells broken down into smaller and smaller time steps surrounding each particle in order to break down the interaction potential. This simulates a version of the Wisdom-Holman symplectic integrator \citep{WisdomHolman1991,Silburt2016}. Comparisons between the {\it Rebound}, MERCURY, and SyMBA hybrid integrators by \cite{Silburt2016} show that although {\it Rebound} took longer to run than SyMBA, it maintained the lowest errors in energy while SyMBA’s energy error grew rapidly. \cite{Rein2017} does a thorough analysis of the {\it Rebound} WHFast, IAS15, and HERMES integrators and concludes that overall {\it Rebound} performs more accurately and at a lower performance cost.



\subsection{{\it Rebound} benefits}
As mentioned, {\it Rebound} \citep{Rein2015} offers the ability to calculate individual collisions between particles due to the N-body nature of the code. Hydrodynamic models struggle to efficiently include the effects of particle collisions since they are not comprised of individual particles.

Additionally, Rebound's structure allows us to add physical effects as individual functions to better simulate specific ejecta scenarios. This creates a flexibility of RED that allows users to select which effects (see table \ref{effectstable}) to include in order to simulate variations on a single impact or to simplify simulations to increase computational efficiency. Some scenarios would be greatly slowed by the addition of effects that have little influence on their outcomes; in these cases, the removal of such effects can improve calculation time without sacrificing accuracy. RED not only uses {\it Rebound} to build simulations, but also is a package of various effects to be implemented into {\it Rebound} for use with other projects outside of the ejecta dynamics application. Ultimately, RED is a Python package of various effects in the solar system that can improve the accuracy of any {\it Rebound} simulation; however, here we primarily focus on the implementation of RED and its application as an ejecta dynamics model using the DART impact as a benchmark system.


\begin{table}
\centering
\caption{List of effects implemented into this study of ejecta dynamics. The effects are listed in the order that they are implemented into the {\it Rebound} simulation. See figure \ref{effectsdiag} for a diagram of these effects. *Future work due to need to first establish basic particle activity in a system. Also, particle-particle collisions follow particle-in-a-box style interactions which differs from other effects implemented here.}
\label{effectstable}
\begin{tabular}{|l|l|}
\hline
\multicolumn{1}{|c|}{\textbf{Order}} & \multicolumn{1}{c|}{\textbf{Effect}}                                                                   \\ \hline
1                                    & Develop basic model of particles being ejected \\ \hline
2                                    & Determine size distribution of particles                                                               \\ \hline
3                                    & Implement non-axisymmetric gravity and rotation                                                              \\ \hline
4                                    & Implement radiation pressure effects \\
                                     & (dependent on particle size \& material)  \\ \hline
5                                    & Apply to binary/triple systems   \\ \hline
6*                                   & Allow particle-particle interactions                                \\ \hline
\end{tabular}
\end{table}

\begin{figure}
    \centering
    \includegraphics[width=.47\textwidth]{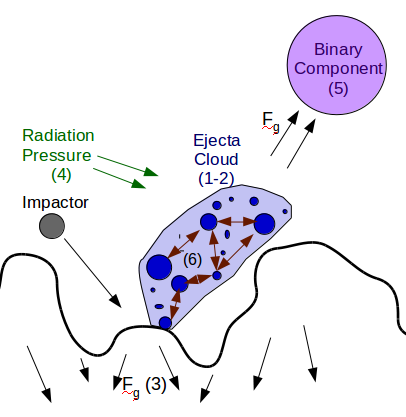}
    \caption{A diagram of the effects implemented in this study. Arrows represent the directions of forces. Numbers next to the different effects correspond to the order in which the effects are implemented. See table \ref{effectstable} for descriptions of each effect.}
    \label{effectsdiag}
\end{figure}

Advances in numerical implementations and computational tools have improved the accuracy of ejecta dynamics models. \cite{Schwartz2016} compares analytic scaling laws, hydrocodes, and N-body integrators in order to determine the most efficient model for excavation of ejecta. This model uses the N-body integrator pkdgrav \citep{Stadel2001} to trace particle dynamics post-excavation. Many previous models do not consider additional effects such as solar radiation and planetary perturbations. Our approach examines a wide array of effects outlined in table \ref{effectstable}. Figure \ref{effectsdiag} shows a sketch diagram of how these effects interact with each other and affect the particles' motions.

\subsection{DART as a Computational Benchmark}
The AIDA/DART mission \citep{Stickle2016,Schwartz2016} to impact a satellite on the binary moon of the Didymos system will help model ejecta distributions in a low gravity vacuum. Due to the close proximity of the Didymos system to the Earth and sun ($a=1.64$AU), the ejecta will be influenced by radiation pressure as well as gravitational effects due to perturbing planets and the non-spherical nature of Didymos and its moon. Initial numerical and computational modeling has been performed \citep{Schwartz2016}; however, more extensive models implementing a wider range of effects -- as we are doing here with RED -- will provide more accurate representations of the expected ejecta plume. These models can be used as predictive tools to determine the trajectories of post-impact debris clouds.

We are applying RED to the DART mission in support of preliminary impact modeling. The Didymos system can be used initially to benchmark RED and refine the added effects. Later, this package will produce a library of possible solutions given a set of initial conditions. Since we will not know the exact initial conditions of the impact until hours before impact, it would be impossible to run a full simulation with the exact set of initial impact conditions. Instead, once the impact conditions are known, we search the library for the solution produced by that particular set of initial conditions. This solution can then be compared to ground and cubesat observations, and we are able to determine the momentum transfer parameter, $\beta$, associated with that particular impact and period change. DART offers the opportunity to examine a real example of an impact on an asteroid to which we can apply RED to learn more about impact conditions on asteroids as well as determine the effects that most heavily influence different types of impacts. The N-body aspect of RED allows us to follow the debris cloud evolution on a small, individual particle scale as well as a larger, system-wide scale.

In this paper we apply the N-body code {\it Rebound} as we implement it as RED to examine the evolution of ejecta at short timescales no longer than a day or two. First, we describe how we implemented {\it Rebound} and six physical effects to study ejecta evolution. Then, we demonstrate several scenarios with the application of the physical effects. Finally, we discuss applications of this code and future developments.

\section{Methodology} \label{sec:methods}

In this section we outline the set-up of our system as well as describe the various effects in the following subsections. Specific details of how each effect was implemented using {\it Rebound} and derivations can be found in Appendix \ref{appendixA}. 

\begin{figure}
{\includegraphics[width=0.47\textwidth]{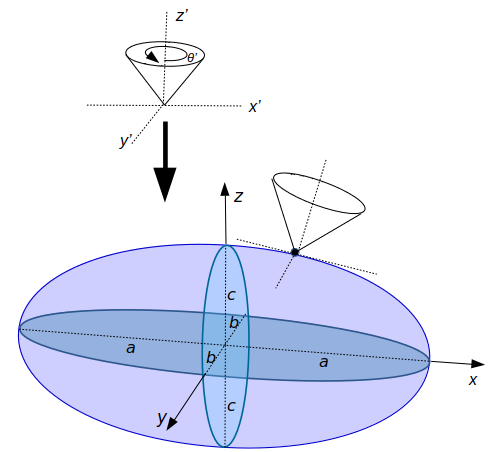}}
{\caption{The target body is placed at the origin of the system, which we define as the universal frame. A secondary local coordinate frame defines particle locations relative to the surface of the target body. For an ellipsoid, we initially align the semi-major axis, \textbf{a}, with the universal $x$-axis. The ejecta cone is initially set up in a local, crater-centric frame before the particles' position and velocity vectors are converted to the universal frame to be added into the {\it Rebound} simulation.}
\label{fig:coordsdiag}}
\end{figure}

Before adding any additional effects or even setting up the system, we must define the coordinate frames used to describe particle positions and motion. The primary coordinate frame can be thought of as the universal frame. This is essentially the grid upon which everything in our system exists, and {\it Rebound} automatically initiates this coordinate frame when a simulation is first set up. It also simplifies calculations to input the target body at the origin of this system such that the semi-major axis, \textbf{a}, lies along the x-axis, \textbf{b} is along the y-axis, and the semi-minor axis, \textbf{c}, is along the z-axis (as seen in \ref{fig:coordsdiag}. All subsequent object positions, forces, and velocities are always input to and output from {\it Rebound} in terms of this universal coordinate frame; however, we are allowed to translate the universal frame into something a bit more comprehensible while calculating how the particles interact with the target body.

Since it is rare for impact scaling laws or observations to have data relative to only the very center of the target, we define a secondary, crater-centric coordinate frame that focuses on the particles' positions relative to the initial impact site. This is primarily required for non-spherical, non-rotating target bodies since the surface features change over time in these cases and the forces experienced by the particles may change based on the positions of the particles relative to the surface. Similarly, mapping particles' positions relative to the impact site (especially when the impact site is on a rotating body) has more meaning to us than simply recording particle positions relative to the center of the target body. 

To convert from the universal coordinate frame to this crater-centric coordinate frame, rather than moving the target itself, we create the illusion of the target moving by moving the universe around the fixed target. For example, a target rotating around the semi-minor axis, \textbf{c}, requires calculating how much the target would have rotated by that time step, then rotating the positions of all the particles clockwise by the rotation amount. From this position relative to the surface, the particles will experience the same forces as it would have if the target body had rotated counter-clockwise and the particles had remained still. While the simplest solution seems to be to simply move the target body relative to the particles, this is in fact far more complicated and would require rotating and manipulating an entire gravitational potential when all that is required is simply where the particles are relative to the surface. Similarly, and this is most important with binary systems, if the ellipsoidal target is tilted with respect to the binary, both the particle positions as well as the binary's position must be rotated with respect to the surface at each time step to account for the system being aligned with a universal coordinate frame. Therefore, at each time step it is absolutely essential to make sure the particles are correctly represented relative to the target's surface.

The initial set-up of particles also requires a coordinate frame shift. Since {\it Rebound} only accepts positions and velocities in terms of the universal frame, we must set up the initial cone in a very local frame in which the particles make up a disk on the x-y plane that covers the entire transient crater. Therefore, to define the size of the initial disk, we use the scaling laws by \cite{Richardson2007} for the volume of a crater to approximate the transient crater radius:

\begin{equation}
    \centering
    V = \frac{1}{3} \pi r_{tc}^{3}
    \label{eq:tcvol}
\end{equation}

\noindent
where $V$ is the transient crater volume and  $r_{tc}$ is the radius of the transient crater. Here we make the same approximation as \cite{Richardson2007} and assume that the crater depth, $H$, is approximately $\frac{1}{3}$ the diameter of the crater, which has been shown in impact experiments to be a reasonable approximation \citep{Melosh1989,schmidthousen1987}. We then calculate the volume of the transient crater by in terms of the material properties of the target ($t$) and the impactor ($i$)\citep{Richardson2007}:

\begin{equation}
    \centering
    V = K_{1} \left( \frac{m_{i}}{\rho_{t}} \right) \left[ \left(\frac{g a}{v_{i}^{2}} \right) \left( \frac{\rho_{t}}{\rho_{i}} \right)^{-\frac{1}{3}} + \left( \frac{\bar{Y}}{\rho_{t} v_{i}^{2}} \right)^{\frac{2+\mu}{2}} \right]^{-\frac{3\mu}{2+\mu}}
    \label{eq:tcvol2}
\end{equation}

\noindent
where $K_{1}$, $\mu$, and $\bar{Y}$ are properties of the target material. Since we assume a target strength of zero, the second term in this equation will also go to zero. By assuming a set of material properties (given by \cite{Holsapple1993}) and impactor parameters (in this study we chose impactor parameters similar to the DART impact), we set \ref{eq:tcvol} and \ref{eq:tcvol2} equal to each other and solve for the transient crater radius, which we will define as the initial particle disk radius: 

\begin{equation}
    \centering
    r_{tc} = \left( \frac{3 V}{\pi} \right)^{\frac{1}{3}} = \left( K_{1} \frac{3 m_{i}}{\pi \rho_{t}} \right)^{\frac{1}{3}} \left[ \left(\frac{g a}{v_{i}^{2}} \right) \left( \frac{\rho_{t}}{\rho_{i}} \right)^{-\frac{1}{3}} \right]^{-\frac{\mu}{2+\mu}}
    \label{eq:tcradius}
\end{equation}

\noindent
This disk lies about 1 cm above the surface to prevent from being counted as particles already landed on the target's surface. 

Additionally, velocities are assigned to each particle based on the \cite{Richardson2007} scaling laws:

\begin{equation}
    \centering
    v_{p} (r) = \left[ v_{e}^{2} - C_{vpg}^{2} g r - C_{vps}^{2} \frac{\bar{Y}}{\rho_{t}} \right]^{\frac{1}{2}}
    \label{eq:veldist}
\end{equation}

\noindent
where $v_{e}$ is the effective velocity that does not go to zero at the transient crater radius \citep{Housen1983}, $\bar{Y}$ is the target material strength, $\rho_{t}$ is the density of the target body, and $C_{vpg}$ and $C_{vps}$ are constants for the gravity dominant regime and the strength regime respectively. Here we only consider the gravity dominated regime and assume a material strength of zero. Therefore, we only need to focus on defining $C_{vpg}$ since the term including $C_{vps}$ goes to zero \citep{Richardson2007}:

\begin{equation}
    C_{vpg} = \frac{\sqrt{2}}{C_{Tg}} \left( \frac{\mu}{\mu + 1} \right)
\end{equation}

\noindent
Note that constant $C_{Tg}$ can be approximated to be $K_{Tg}$. Both $K_{Tg}$ and $\mu$ are properties of the target material specified by \cite{Holsapple1993}; here we assume the target to be made of a material similar to sand so $K_{Tg}=0.5$ \citep{Housen2011} and $\mu = 0.41$ \citep{Holsapple1993}. 

The core component of the velocity distribution is the effective velocity equation by \cite{Housen1983}, which was later improved upon by \cite{Richardson2007} to include the gravity and strength terms. This component is defined by \cite{Richardson2007} as

\begin{equation}
    v_{e} (r)= C_{vpg} \sqrt{g R_{g}} \left( \frac{r}{R_{g}} \right) ^{-\frac{1}{\mu}}
    \label{eq:veff}
\end{equation}

\noindent
where $R_{g}$ is the transient crater radius and $r$ is the radius at which the particle is initiated (this is between zero and $R_{g}$). We substitute equation \ref{eq:veff} into equation \ref{eq:veldist} to determine a velocity distribution based on the material properties and impact parameters.

\begin{figure}
    \includegraphics[width=.47\textwidth]{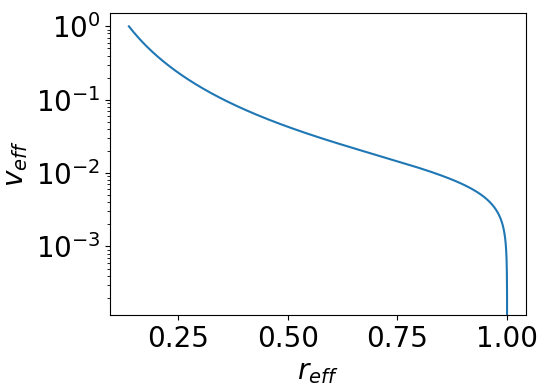}
    \caption{Normalized velocity distribution of particles based on radial distance from the center of the crater to the transient crater radius. Particles closer to the center of the ejecta plume exit the surface at a higher velocity than particles towards the outside of the plume. This is particular scaling example is calculated based on an impact velocity of $6.6$ km $\mathrm{s^{-1}}$, which is the projected impact velocity for the DART impact \citep{cheng2016}}
    \label{fig:veldist}
\end{figure}

A version of this velocity distribution can be seen in figure \ref{fig:veldist}, which depicts the effective velocity at different effective crater radii. The effective velocity and effective radius here are the velocities and radii for an impact similar to what will be seen for the DART impact divided by the highest velocity and the transient crater radius respectively. All particles are ejected at an angle of $45^{\circ}$ relative to the local surface plane. While ejection angles tend to vary slightly according to experiments by \cite{Cintala1999}, we simply chose to have a uniform ejection angle with varying velocities. Varying the initial ejection angle did not have a significant impact on the overall long-term evolution of the ejecta in our simulations.

After assigning positions and velocities to the particles in the local crater frame, we translate the vectors into the universal frame to be added to the {\it Rebound} simulation. We use a 3D rotation matrix and the impact latitude and longitude to rotate the particle vectors from the crater frame to the universal frame. Since this conversion between the local frame and universal frame typically occurs at least once every time step in order to recalculate forces and additional effects, we include a specific 3D rotation matrix function in the python package for {\it Rebound}.

\subsection{Gravitational Effects}
\label{ss:graveff}

Our initial model assumes all bodies are spherical. However, shape models of asteroids and other small bodies show that small bodies are not typically perfectly spherical; rather, they exhibit more ellipsoidal or "potato" shapes. Particles near the surface of an ellipsoidal body will experience a different gravitational force depending on their location relative to the body. Therefore, we implement an ellipsoidal gravitational potential for the target body as well as any potential secondary components. While other methods for calculating irregular gravitational fields exist (e.g. \cite{werner1994}), we chose the ellipsoidal gravitational potential as a way to minimize computational requirements while still maintaining a more accurate approximation than a sphere. In future development stages of the RED package, we plan to include more advanced methods of gravitational filed mapping so that we can consider more complex shape models and examine the more subtle physical effects that may arise in such situations. We will implement it either through additional numerical modules or ingested look-up tables with interpolation. In general, complex shape models can be based on more detailed analysis and interpretation, such as those derived from data acquired during in-situ robotic missions to small bodies, or ground-based radar observations.

Next, we include the rotation of the target body as an additional velocity vector added to the particles near the target body. In order to produce this effect, we calculate the linear velocity of the impact point on the surface at time zero. For a body with a vertical axis of rotation, this will translate to a velocity vector in the x and y directions, but if the axis of rotation is tilted, a $z$ component will be included in the vector. Similarly to how we set up the initial ejecta distribution, we begin by assuming that the axis of rotation is vertical along the z-axis. Then, we calculate the instantaneous velocity at time zero for the latitude and longitude of the impact site. Finally, the 3D rotation matrix function tilts the instantaneous velocity vector the amount that the rotation axis is tilted. Note: the direction in which the axis is tilted must be specified and included in the rotation matrix, and is given by some angle between $0^{\circ}$ and $360^{\circ}$ in the xy-plane. Due to the nature of how this code is set up, no additional accelerations need to be added to the particles. As the particles leave the surface of the body, they will also have an added velocity due to the instantaneous rotation velocity at the point at which the particles left the body. Once the particles are off the surface, they are only affected by the gravitational field (and any radiation pressure effects) and no longer influenced directly by the rotation of the target body.

Finally, including other bodies orbiting the target body introduces an additional gravitational acceleration on the particles. Orbital parameters, mass, and radius of the binary are input in relation to the target body; therefore, if two bodies of similar mass orbit their center of mass, we will observe this system from the perspective of one of the bodies rather than from the perspective of the center of mass. When this secondary body is added to the system, it is placed in orbit around the target body.

\subsection{Effects Dependent on Size Distribution}
In addition to gravitational effects, we include effects dependent on particle size, such as radiation pressure. We use a power law distribution similar to \cite{OBrian2003} to determine the radius of each particle before they are added to the simulation:

\begin{equation}
\centering
dN = C R^{-p} dr_{p}
\label{powerlaweq}
\end{equation}

\noindent
where $C$ is a constant, $N$ is the number of occurrences of a particle with radius $r_{p}$ and $p$ is the power of the distribution. In our simulations we chose to use $p = 3$, which is consistent with other size distributions where $2 < p < 4$ \citep{OBrian2003}. By integrating equation \ref{powerlaweq} between chosen minimum and maximum radii, we determine the number of particles for each radius. Since equation \ref{powerlaweq} only provides the continuous solution, a certain resolution between the minimum and maximum radii must be selected. The resolution describes the number of desired radii between the minimum and maximum radii, providing an incremental solution that can be applied to the particles themselves. For these simulations, we use a range of radii $10^{-4}\text{cm} < r_{p} < 1\text{cm}$ and a resolution of 100. Each particle is then assigned to a bin corresponding to a certain radius. Equation \ref{powerlaweq} defines how many particles can be assigned to each bin. We chose to use 100 bins for these simulations because for a simulation of $10^{4}$ particles, using more bins makes it more difficult to distinguish between distinct particle size groups and the effects on these groups, and using fewer bins leads to too low a resolution to see distinct effects.

\begin{figure}
\centering
\caption{The ratio of acceleration due to radiation pressure to acceleration due to gravity for varying particle sizes being ejected from a 100m radius circular asteroid. Each line represents the ratio at a certain distance from the surface of the body, the bottom line representing 0m from the surface and the top line indicating 400m from the surface.}
\label{fig:radpressparts}
\includegraphics[width=0.47\textwidth]{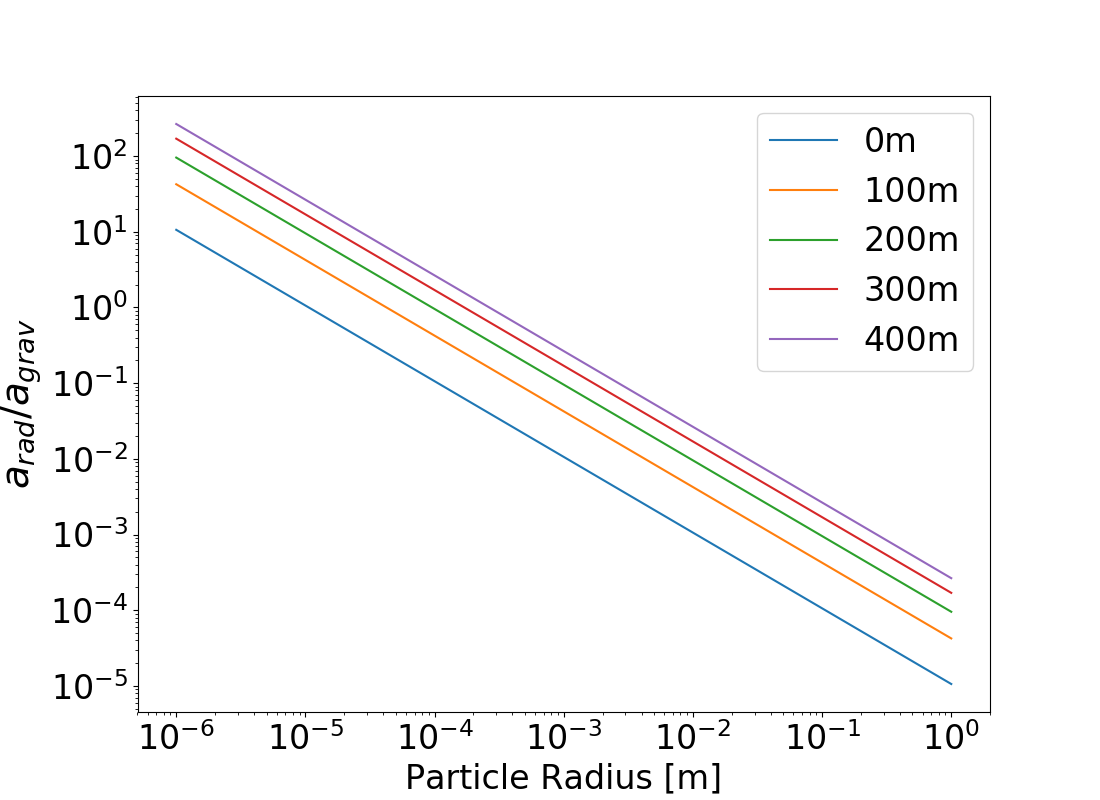}
\end{figure}

While the particle sizes do not have a large effect on gravitational forces, radiation pressure from the sun exerts a slight but perceivable force on smaller ejecta particles than the larger particles. Figure \ref{fig:radpressparts} shows the influence of the radiation pressure force relative to the gravitational acceleration felt by particles of varying radii. This effect is relative to the distance from the sun as well as the radius of the particle:

\begin{equation}
    \centering
    a = \frac{3 \epsilon K_{sc}}{4 r \rho c}
    \label{eq:aPR}
\end{equation}

\noindent
where $K_{sc}$ is the solar constant calculated at the target body's distance from the sun, $r$ is the particle radius, $c$ is the speed of light, and $\epsilon$ is an absorption factor. For our purposes, we assign $\epsilon$ to be 1.5 as an average value for all particles. Derivations for equation \ref{eq:aPR} can be found in Appendix \ref{appendixA}. We also include a shadowing effect so that as particles pass "behind" a large body in the system, the radiation pressure goes zero. For this addition, we assign a cylinder with the radius of the object and assume that this cylinder goes to infinity from the object in the opposite direction of the sun. Any particle within this cylinder does not experience any additional radiation pressure forces.

The Cartesian unit vector of this acceleration is determined based on the unit vector from the sun to each individual particle. Finally the acceleration components can be added to the acceleration components for each of the particles. This force creates a drag on the particles that is proportional to the size of the particle. Similarly, particles beyond 2.5 AU from the sun feel little to no radiation effects from the sun. Therefore, in simulations that take place beyond 2.5 AU can neglect all radiation effects.
\begin{figure}
\label{FvsR}
\end{figure}

\section{Results} \label{sec:results}
To test the effects described in the previous section, we simulate a cloud interacting with a binary asteroid system (Model A) similar to the Didymos system. Information about the system parameters can be referenced in table \ref{systemparams}. This section discusses the test scenarios used to demonstrate each of the additional effects. Initial set-ups for all simulations are described in table \ref{partstable} where each column corresponds to a figure highlighting the progression of that simulation. 


\begin{table}
\centering
\caption{General parameters for Model A, a binary asteroid system modeled loosely based on the Didymos system. Model A1 represents the primary component and Model A2 represents the secondary component of the binary system. The orbital data for A2 are in reference to its orbit around A1 while the A1 data describe the system's orbit about the sun. Data presented below courtesy of the JPL Horizons Database.}
\label{systemparams}
\begin{tabular}{|l|l|l|l|}
\hline
System                                     & Model A1                 & Model A2             \\ \hline
Mass {[}kg{]}                              & $4.56\times 10^{11}$     & $3 \times 10^{9}$    \\ \hline
Radius {[}m{]}                             & 400                      & 75                   \\ \hline
Binary                                     & Yes                      & Yes                  \\ \hline
Orbital Radius                             & 1.64 AU                  & 1.2 km               \\ \hline
Eccentricity                               & 0.38                     & -                    \\ \hline
Inclination                                & $3.41^{\circ}$           & -                    \\ \hline
Lon. of Asc. Node ($\Omega$)               & $73.21^{\circ}$          & -                    \\ \hline
Arg. of Periapsis ($\omega$)               & $319.30^{\circ}$         & -                    \\ \hline
Esc. Vel. ($v_{esc}$) [m ${\rm s^{-1}}$]   & 0.39                     & .07                  \\ \hline
\end{tabular}
\end{table}

\begin{table}
\centering
\caption{All particles are initiated with the same basic ejecta cone parameters; however, we vary the additional effects applied to each simulation. Figure \ref{fig:noeffD} represents the basic case with no additional effects. Figure \ref{fig:ellipD2} is the case in which the target is ellipsoidal with an axis ratio similar to the estimated ratio for Dimorphos. To demonstrate a larger extend of the ellipsoid's gravitational influence, we chose a longitude of $45^{\circ}$ so that the ejecta would not lie directly on an ellipsoid axis. Figure \ref{fig:rotD2} contains a target body that rotates around a vertical axis once every 6 hours. Figure \ref{fig:binD} contains a binary system where the smaller object, A2, is the target and the larger, A1, orbits around the target. Finally, figure \ref{fig:radD} employs a particle size distribution so that smaller particles are influenced by radiation pressure.}
\label{partstable}
\begin{tabular}{|l|l|l|l|l|l|}
\hline
Figure Number                     & \ref{fig:noeffD}    & \ref{fig:ellipD2} & \ref{fig:rotD2}   & \ref{fig:binD}       & \ref{fig:radD}    \\ \hline
Longitude ($\theta'$)             & $90^{\circ}$        & $45^{\circ}$      & $90^{\circ}$      & $90^{\circ}$         & $90^{\circ}$      \\ \hline
Latitude ($\phi'$)                & $0^{\circ}$         & $0^{\circ}$       & $0^{\circ}$       & $0^{\circ}$          & $0^{\circ}$       \\ \hline
Target Axis Ratio                 & -                   & 1.82:1.36:1       & -                 & -                    & -                 \\ \hline
Rotation Per. [hr]              & 0                   & 0                 & 6                 & 0                    & 0                 \\ \hline
Binary?                           & No                  & No                & No                & Yes                  & No                \\ \hline
Particle Radii [m]                & -                   & -                 & -                 & -                    & $10^{-4}$ to 1    \\ \hline

\end{tabular}
\end{table}

\subsection{Gravitational Effects}

\begin{figure}
    \centering
    \begin{subfigure}[b]{\textwidth}
        \includegraphics[width=.47\textwidth]{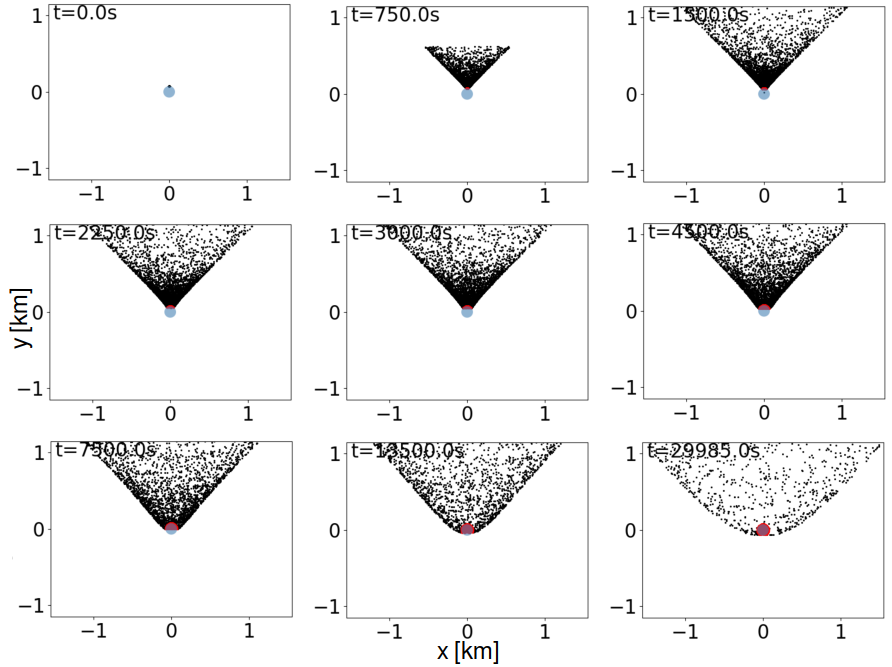}
    \end{subfigure}
    
    \vspace{.05cm}
    
    \begin{subfigure}[b]{\textwidth}
        \includegraphics[width=.47\textwidth]{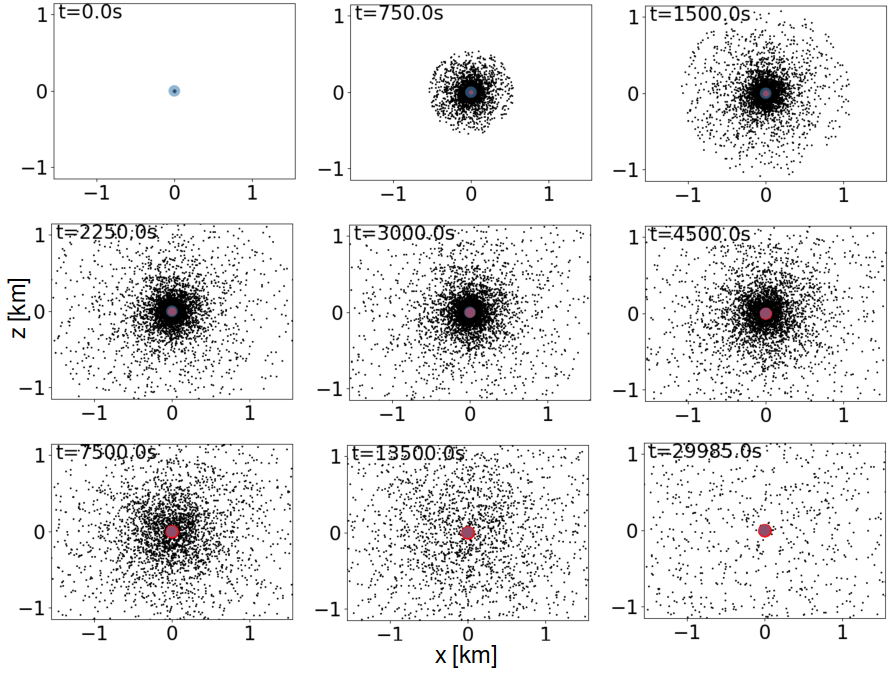}
    \end{subfigure}
    
\caption{Initial test of basic ejecta plume with no additional effects on Model A2. Here we ignore any binary effects by removing A1 from the simulation. The first nine panels show snapshots of the x-y projection while the latter nine panels are snapshots of the x-z projection. Specific details regarding the exact initial parameters of this simulation can be found in table \ref{partstable}. \label{fig:noeffD}}

\end{figure}

First, we start with the simplest case involving no additional effects to ensure that the basic ejection of material is operating correctly before complicating the simulations with more complex effects. As shown in figure \ref{fig:noeffD}, the target body is spherical, so ejecta particles experience a consistent gravitational acceleration from a single body regardless of the initial longitude and latitude of the impact event. High velocity particles near the center of the impact are rapidly lofted far from the surface. The lower velocity particles towards the outside of the crater land close to the edge of the crater while leaving the impact site free of debris. Overall, we observe that $65.37\%$ of particles land on the target body throughout the course of the simulation. Even with only a portion of all $10^{4}$ particles landing back on the body, the entire target body becomes covered in debris, primarily due to the relatively small target body size combined with the high impact velocity of 6.6 km $\mathrm{s^{-1}}$ which allows particles to loft farther from the body. This simulation establishes a baseline against which all subsequent simulations with additional effects can be compared. 


        
    
    



\begin{figure}
    \centering
    \begin{subfigure}[b]{\textwidth}
        \includegraphics[width=.47\textwidth]{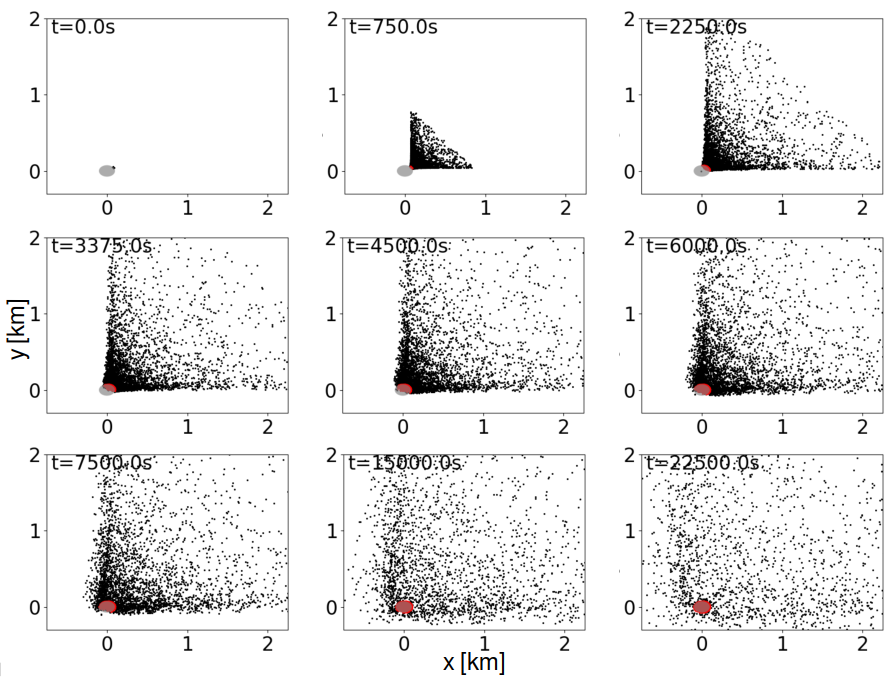}
    \end{subfigure}
    
    \vspace{.05cm}
    
    \begin{subfigure}[b]{\textwidth}
        \includegraphics[width=.47\textwidth]{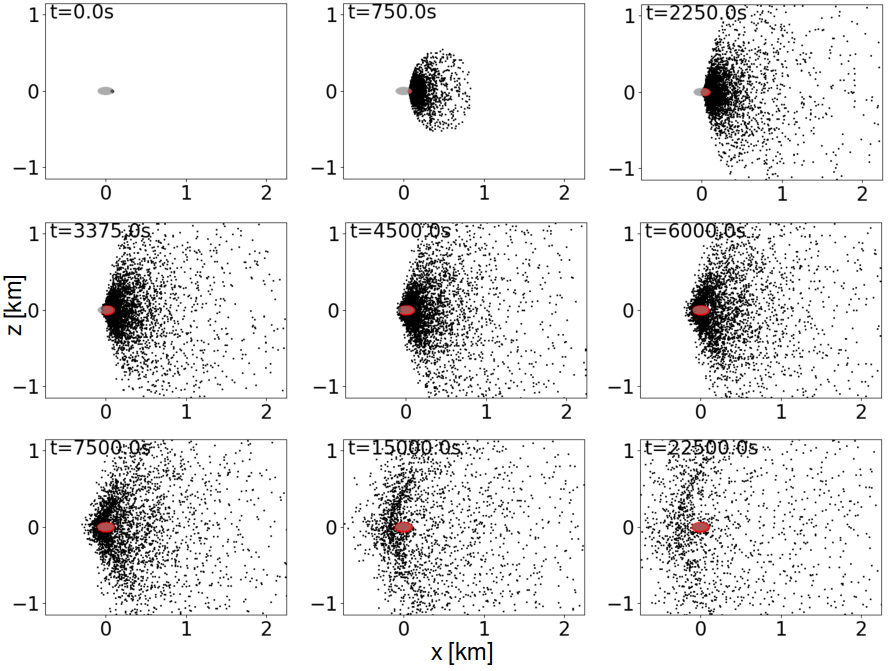}
    \end{subfigure}
    
\caption{A simulation of a cloud off of an ellipsoidal A2. Changing the ejection location on an ellipsoidal body changes the results due to particles being introduced to a completely new gravity field. Here particles are ejected on the equator at a longitude of $45^{\circ}$. The first nine panels show snapshots of the x-y projection while the latter nine panels are snapshots of the x-z projection. Specific details regarding the exact initial parameters of this simulation can be found in table \ref{partstable}.}
\label{fig:ellipD2}
\end{figure}

While in many situations it is realistic to use a spherical target body, often smaller bodies are more ellipsoidal; therefore, it is more reasonable to implement the ellipsoidal gravitational potential described in section \ref{ss:graveff} as an initial step towards more advanced gravitational potential implementations. In figure \ref{fig:ellipD2}, we simulate the drift of particles due to the varying gravitational potential where the initial particle distribution is $45^{\circ}$ from the major axis of the ellipsoid and lies along the equator of the ellipsoid. The longer axis of the ellipsoid has a higher gravitational potential than the other axes of the ellipsoid and can be seen here, causing the cloud to drift towards the lobes with higher potential; however, this effect is slight and only noticeable close to the target surface. Farther from the body, the ellipsoidal potential appears closer to the point mass potential created by the spherical target body. In this scenario, $64.71\%$ of the $10^{4}$ particles landed on the surface. Since this is similar to the amount of particles that landed in the no effects scenario, we determine that the ellipsoidal gravitational potential does not greatly affect particle trajectories in the long-term or the far-field, but the final distribution of landed particles may change due to surface variations.

\begin{figure}
    \centering
    \begin{subfigure}[b]{\textwidth}
        \includegraphics[width=.47\textwidth]{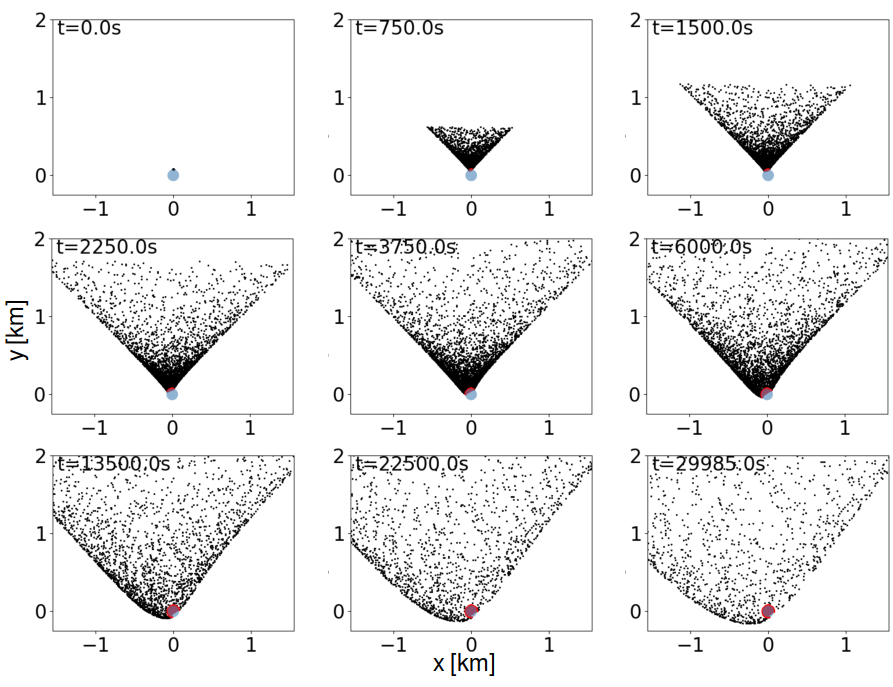}
    \end{subfigure}
    
    \vspace{.05cm}
    
    \begin{subfigure}[b]{\textwidth}
        \includegraphics[width=.47\textwidth]{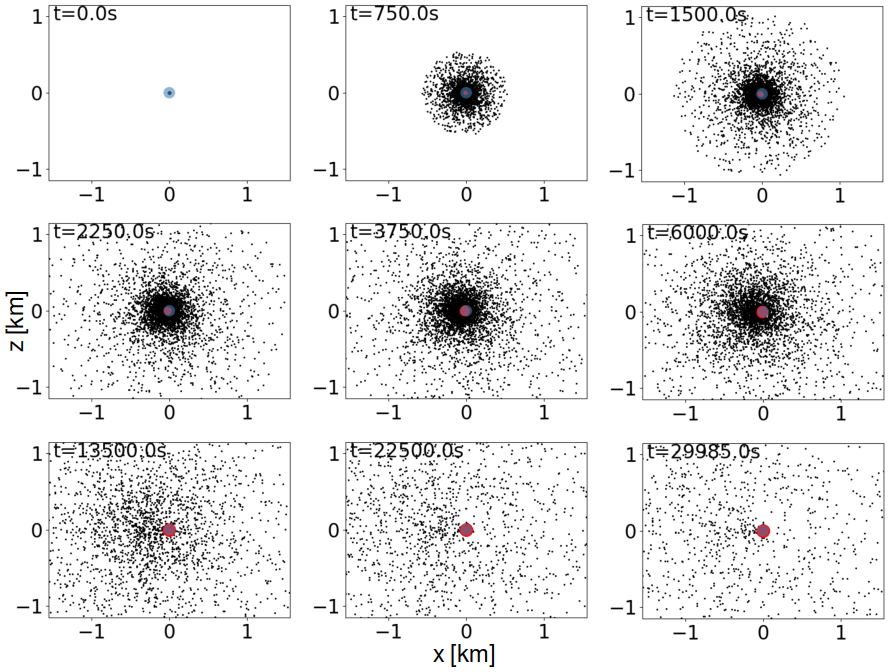}
    \end{subfigure}
    
\caption{Here target body A2 is spherical and rotates once every 6 hours about a vertical axis. Particles are given an additional rotational velocity incorporated into the ejection velocity at the initial set-up. The first nine panels show snapshots of the x-y projection while the latter nine panels are snapshots of the x-z projection. Specific details regarding the exact initial parameters of this simulation can be found in table \ref{partstable}.}
\label{fig:rotD2}
\end{figure}

In addition to an ellipsoidal potential, rotation of the target body alters near-field ejecta dynamics. As particles exit the surface, they retain the linear velocity caused by the rotation of A2 at the instant they left the surface, as shown in figure \ref{fig:rotD2}. High velocity particles are lofted at a much faster rate than the rotation velocity, so the ejection velocity component dominates over the rotational component. Slower particles experience a larger influence from the rotational component and therefore appear to continue rotating along with the body above the surface. While rotation does not largely affect the quantity of particles that fall back on the surface (in this scenario $64.26\%$ of all particles landed on A2), the distribution of particles is greatly altered due to material ``drifting" off of the original impact site. In simulations with A2 as the target body, once again the majority of the surface is covered in debris; however, larger bodies or less energetic impact scenarios may lead to a final ejecta distribution that looks similar to the particle distribution caused by an oblique impact on a non-rotating object. Additionally, combining rotation with the ellipsoidal gravitational potential creates a gravitational potential that varies over time at a given point in space.

\begin{figure}
    \centering
    \begin{subfigure}[b]{\textwidth}
        \includegraphics[width=.47\textwidth]{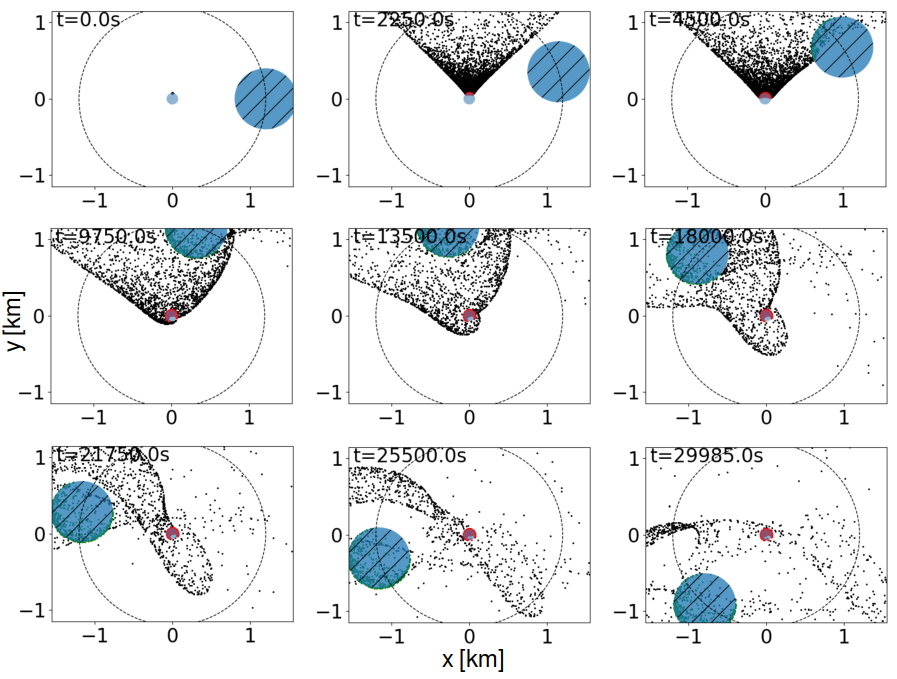}
    \end{subfigure}
    
    \vspace{.05cm}
    
    \begin{subfigure}[b]{\textwidth}
        \includegraphics[width=.47\textwidth]{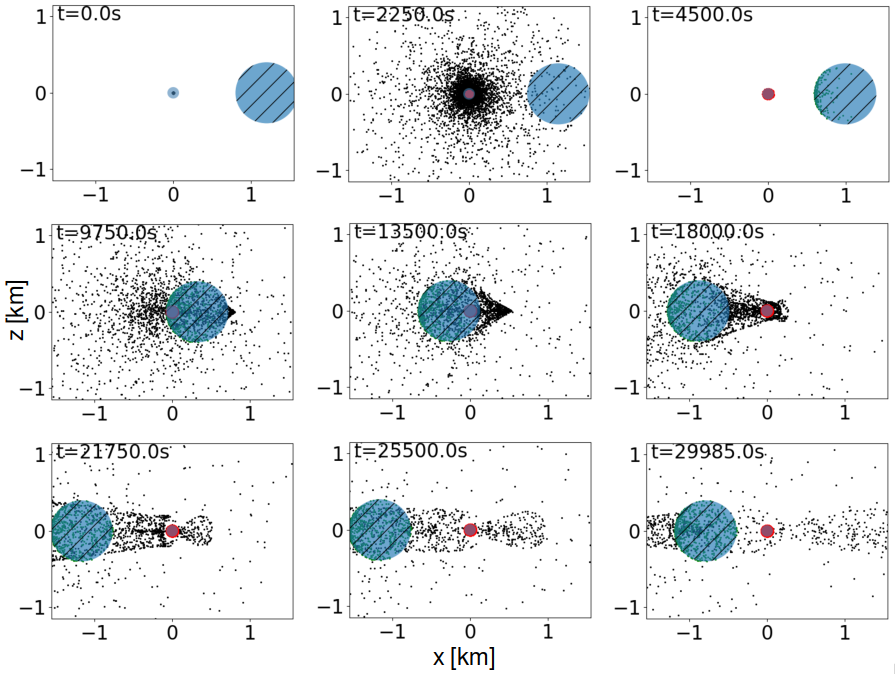}
    \end{subfigure}
\caption{Both A1 (denoted by large circle with hash marks) and A2 (denoted by solid circle) are assumed to be spherical with the ejecta exiting A2 $90^{\circ}$ from A1. The first nine panels show snapshots of the x-y projection while the latter nine panels are snapshots of the x-z projection. Specific details regarding the exact initial parameters of this simulation can be found in table \ref{partstable}.}
\label{fig:binD}
\end{figure}

\begin{figure}
    \centering
    \begin{subfigure}[b]{.75\textwidth}
        \includegraphics[width=.55\textwidth]{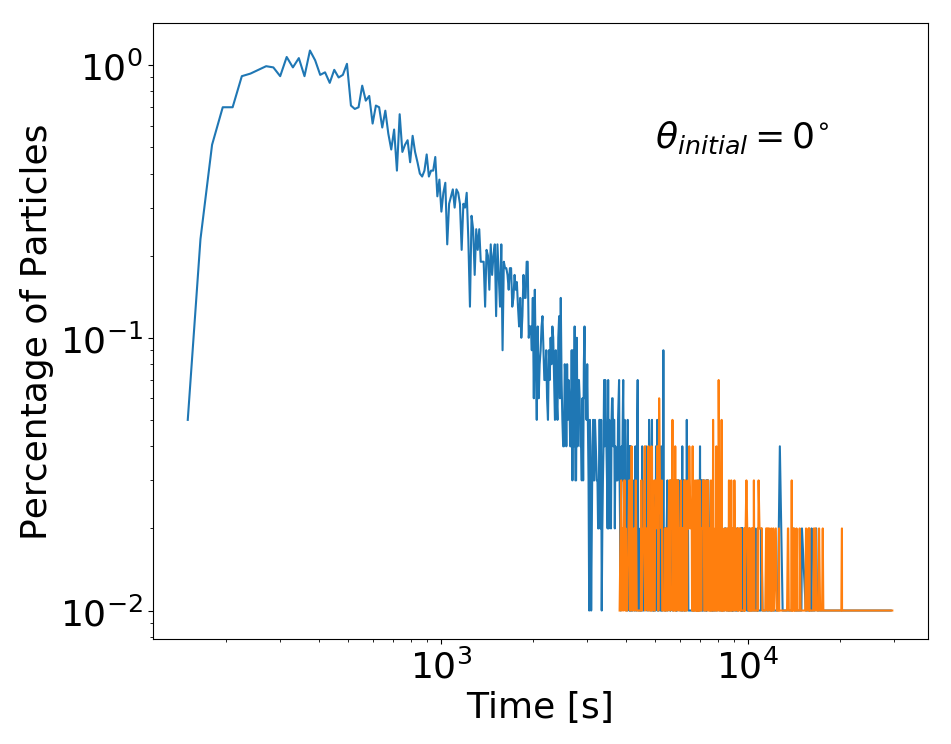}
        
    \end{subfigure}
    
    \begin{subfigure}[b]{.75\textwidth}
        \includegraphics[width=.55\textwidth]{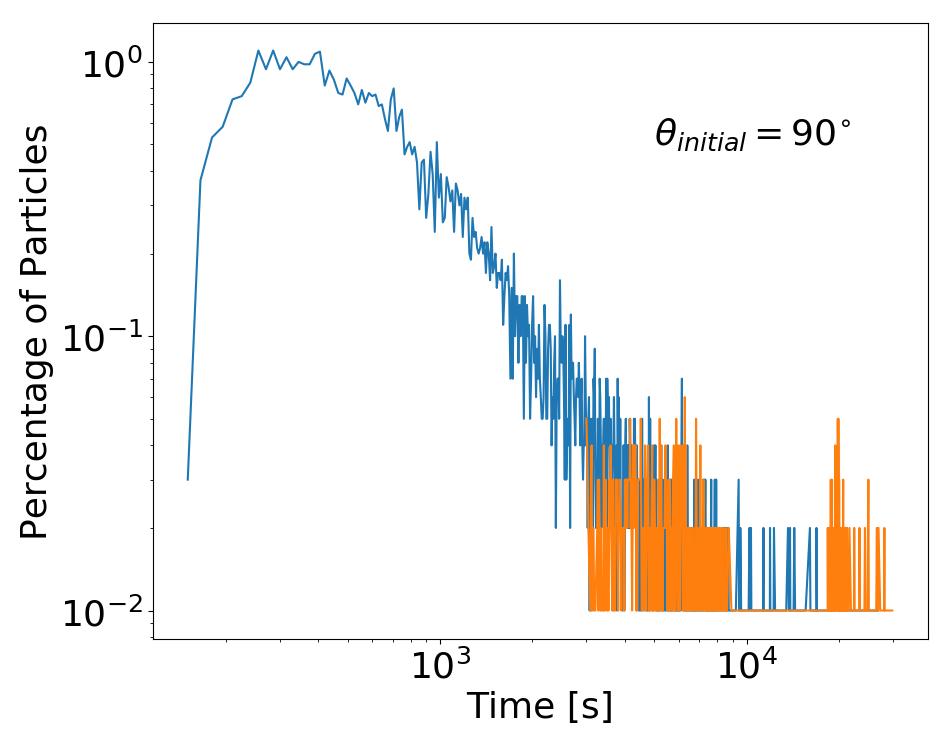}
        \label{subfig:binnum2}
    \end{subfigure}
    
    \begin{subfigure}[b]{.75\textwidth}
        \includegraphics[width=.55\textwidth]{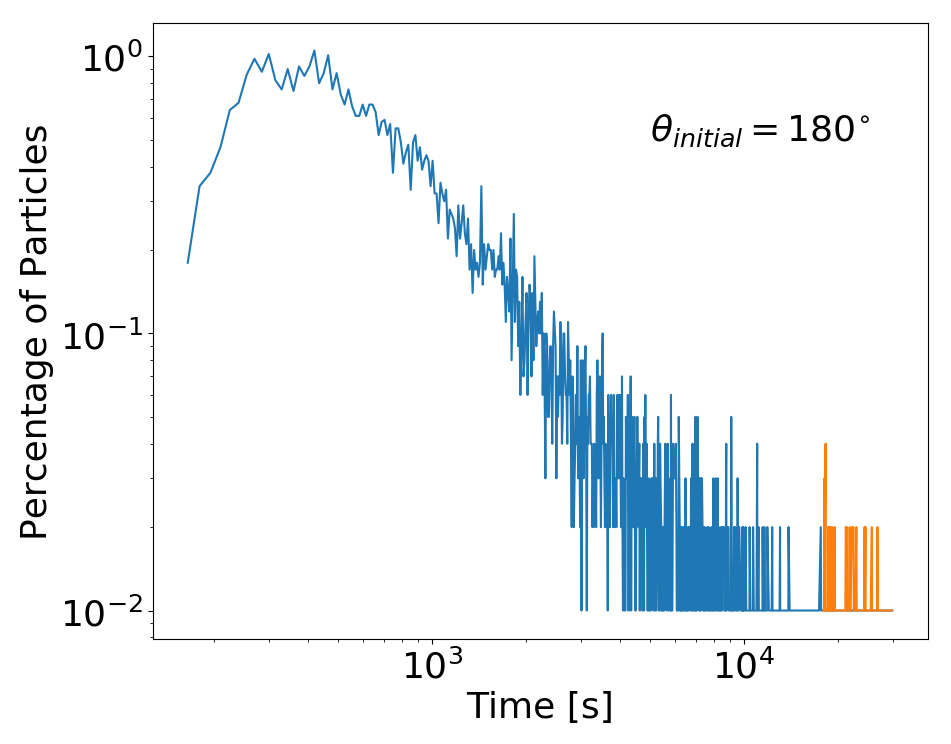}
        \label{subfig:binnum3}
    \end{subfigure}
    
    \begin{subfigure}[b]{.75\textwidth}
        \includegraphics[width=.55\textwidth]{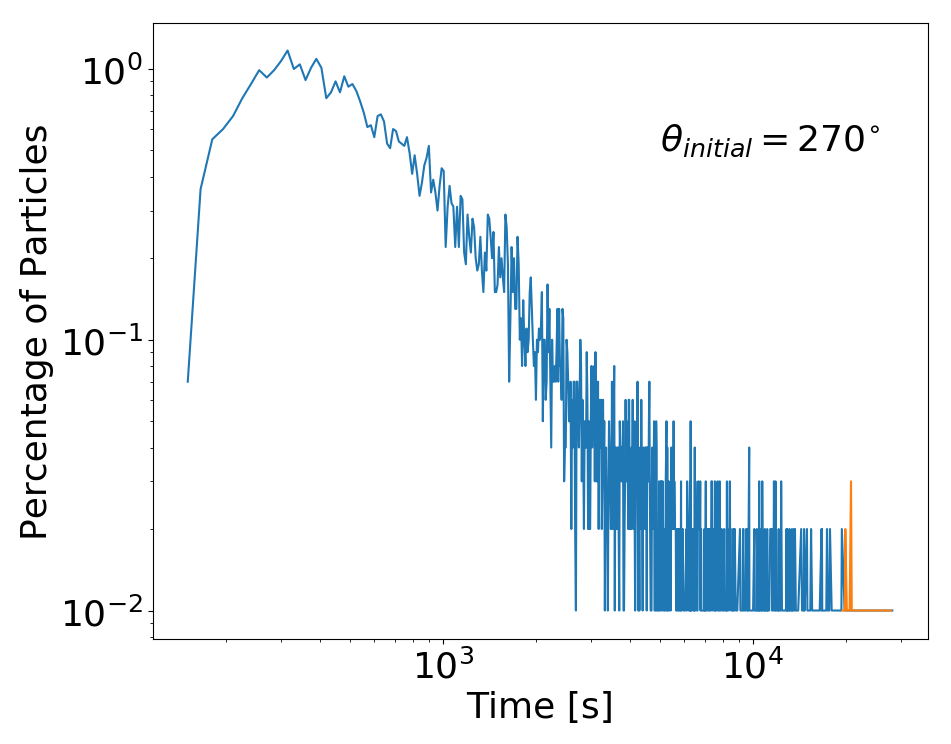}
        \label{subfig:binnum4}
    \end{subfigure}
    \caption{A certain percentage of particles in a binary system fall on the orbiting body (A1) while the rest fall back on the target body (A2). The darker line depicts the percentage of particles that fall back onto A2 at each time step, and the lighter line represents the percentage of particles that fall back on A1. From top to bottom, the starting locations occur at the following longitudes: $0^{\circ}$, $90^{\circ}$, $180^{\circ}$, and $270^{\circ}$}
    \label{fig:binnum}
\end{figure}

We designed the Model A system as a binary asteroid system in order to test the effects of a secondary component on an ejecta plume. Post-impact debris clouds in binary systems not only experience a gravitational influence from the target body, but they also are influenced by the secondary component. Particles that pass beyond the Roche limit of the target body leave the gravitational influence of the target body and may be influenced by the gravity of the secondary object, as seen in figure \ref{fig:binD}. In binary systems, the orientation of the impact with respect to the secondary object plays a crucial role in the final distribution of particles. Figure \ref{fig:binD} demonstrates an impact on A2 $90^{\circ}$ from the position of body A1, where A1 orbits counterclockwise towards the impact site. Object A1 sweeps up the particles from the ejecta cone and pulls the particles around A2. Similar interactions occurred for the other three binary scenarios in which we altered the location of the impact site relative to A1, which lies on the positive x-axis. As seen in table \ref{binperc} and figure \ref{fig:binnum} (a graphical representation of table \ref{binperc}), roughly the same amount of particles land on A2 regardless of the starting position; however, more particles land on A1 if the starting position near the start of the orbit (between $0^{\circ}$ and $90^{\circ}$). Particles in the $0^{\circ}$ and the $90^{\circ}$ longitude scenarios eject particles directly onto A1 or immediately into the orbit of A1 to be swept up and do not have as much to to disperse or fall back on A2 as in the $180^{\circ}$ and $270^{\circ}$ scenarios. Based on figure \ref{fig:binnum}, most of the material landing on A1 has already done so by the time particles begin landing on A2, so there are less particles that have the chance of falling onto A2. Comparing to the no effects simulation, we note that the presence of a binary component as well as the location of the impact event relative to the binary component significantly affect the evolution of the ejecta plume.

\begin{table}
\centering
\caption{Percentages of particles that landed on A1 and A2 in the binary system simulations depicted in \ref{fig:binD} (for the $90^{\circ}$ case) and \ref{fig:binnum}. The impact site angle on A2 is defined by looking at A2 from above (cross-section of the xy-plane) where the positive x-axis is $0^{\circ}$ and the positive y-axis is $90^{\circ}$. For each scenario A1 begins at $0^{\circ}$ and orbits counterclockwise.}
\label{binperc}
\begin{tabular}{|l|l|l|}
\hline
Impact Site on A2     & $\%$ on A1             & $\%$ on A2            \\ \hline
$0^{\circ}$           & 7.41                   & 64.36                 \\ \hline
$90^{\circ}$          & 7.78                   & 64.80                 \\ \hline
$180^{\circ}$         & 1.68                   & 63.57                 \\ \hline
$270^{\circ}$         & 0.60                   & 64.95                 \\ \hline
\end{tabular}
\end{table}


\subsection{Effects Dependent on Size Distribution}
\begin{figure}
    \centering
    \begin{subfigure}[b]{\textwidth}
        \includegraphics[width=.47\textwidth]{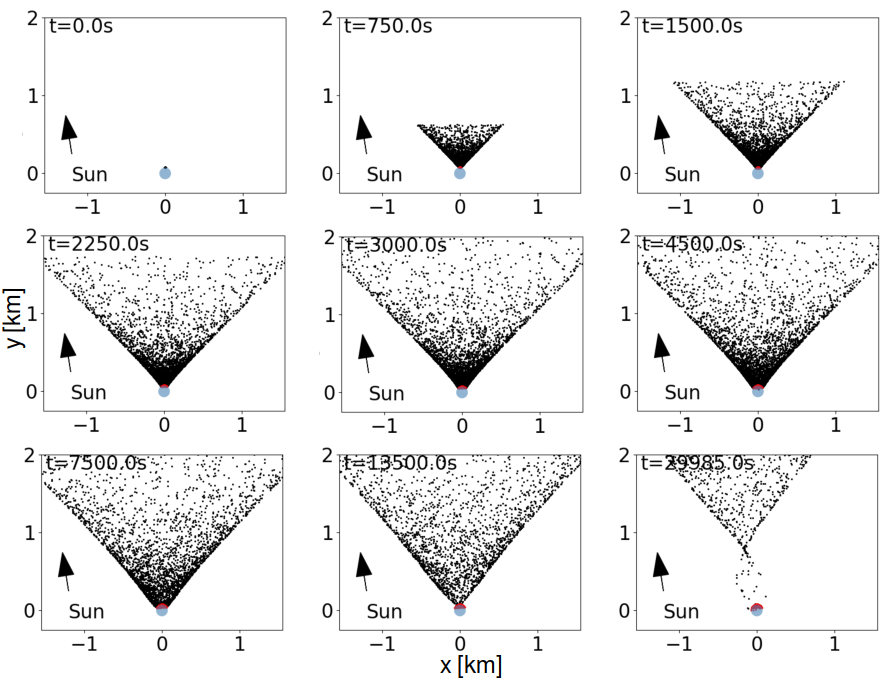}
    \end{subfigure}
    
    \vspace{.05cm}
    
    \begin{subfigure}[b]{\textwidth}
        \includegraphics[width=.47\textwidth]{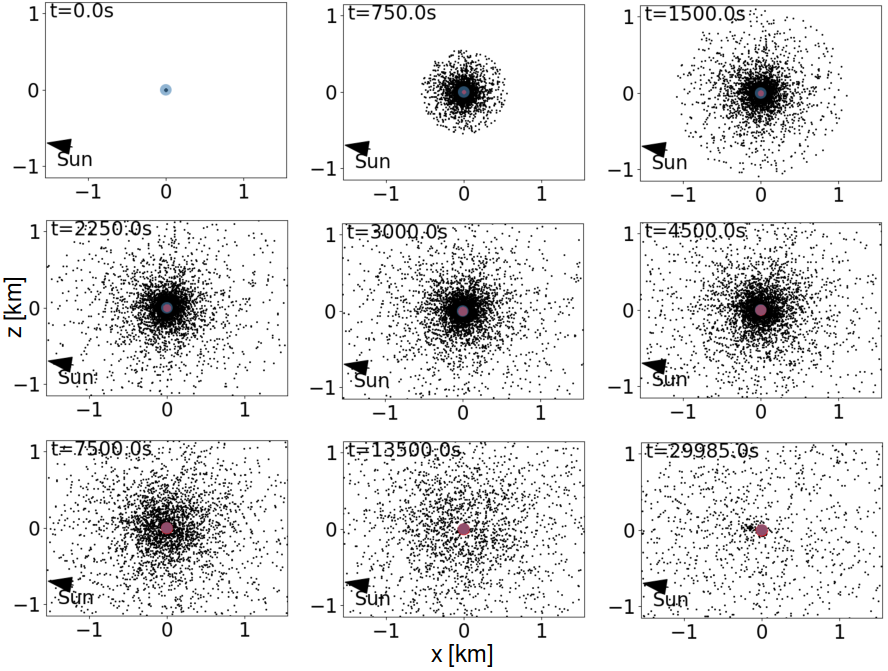}
    \end{subfigure}
\caption{The particle size distribution implemented here only plays a role in the particle trajectories when coupled with solar radiation pressure. We assume A2 to be spherical. The arrow labeled "Sun" points away from the sun, denoting the acceleration vector that the particles experience due to radiation pressure. The first nine panels show snapshots of the x-y projection while the latter nine panels are snapshots of the x-z projection. Specific details regarding the exact initial parameters of this simulation can be found in table \ref{partstable}.}
\label{fig:radD}
\end{figure}

Radiation pressure mostly affects the smaller particles in an ejecta plume; similarly, the target body must have an orbit close enough to the sun to experience the solar radiation pressure. Our Model B asteroid system can be classified as a Near Earth Object (NEO), orbiting the sun at approximately 1.6 AU, close enough to the sun to experience a fairly strong radiation pressure in comparison to objects with orbits outside 2.5 AU. Figure \ref{fig:radD} provides an example of the particle drift due to radiation pressure pushing particles away from the sun. Here the particles are initially ejected on the side of A2 opposite the sun. Over time the particles drift slightly to align with this vector moving away from the sun. Particles behind A2 fall into its shadow where they experience no radiation pressure, so at first the radiation effects are not seen. However, as the particles leave the shadow, the ejecta cone narrows and drifts away from A2. While $62.58\%$ of particles fall back on the surface (just slightly under the amount that fall back in the no effects scenario), only the shadowed side of A2 accumulates debris compared to the entire surface when there are no effects. Ejecta that normally would have drifted around the entire surface are prevented from even crossing over to the sun-ward side of the target. Similar to in the binary system case, the position of the impact site relative to the sun influences the plume evolution and final particle distribution.

The next step we intend to take in the development of this package is to include particle-particle interactions. While not as important in the later stages of ejecta plume evolution, inter-particle collisions play a crucial role in defining the particle size and velocity distributions during the early lofting stages of evolution. We anticipate that depending on the initial velocity regimes defined, the overall distribution will either experience an increase in particle size due to accretion (thus inhibiting radiation pressure effects) or particle size will decrease due to the break up of other particles during collisions (thus increasing radiation pressure effects). Further development of this function is necessary in able to demonstrate accuracy of these anticipated results.

\section{Discussion and Future Work} \label{sec:discussion}


Applications for the RED package range from modeling of ejecta to cometary outbursts to surface activity. As discussed previously, we are applying this model to the preliminary modeling of the DART impact. Since the exact impact parameters are not yet known and will not be known for certain until just before the impact, we are assembling a library of possible outcome solutions based on different impact conditions. Due to the shorter required computational time that {\it Rebound} provides, we are able to produce a large number of simulation outcomes (approximately 200 ejecta scenarios) much faster than other N-body codes can. Each simulation will have some assigned momentum transfer parameter, $\beta$, that corresponds to a set of initial conditions. Observations of the impact provide a final distribution from which we can reverse engineer a set of initial conditions. We can determine the corresponding $\beta$ associated with an impact of similar conditions by comparing the observations to our library of simulations.

In a future companion paper, we will introduce and benchmark the RED package's particle-particle interaction suite. This function will calculate relative velocities of colliding particles. The velocity regime in which the collision resides determines the type of collision that occurs. This can be generally distributed into low, medium and high velocity regime, where the particles can bounce off in an elastic manner, disrupt and re-accumulate, or have a hybrid behavior, which is probably sensitive to specific collision conditions (impact angle, mass ratio, location in the gravitational potential of the asteroid system).  

Additionally, this package can be applied to a wide range of other dynamical systems, such as cometary and Centaur outbursts. While this package does not specifically include dynamical effects from the gas, additional forces may be added to imitate the influence of the gas on the particles. Analyzing simulation results in the form of light curves allows for comparisons to outburst observations. Similarly, due to the bit-wise integration scheme built into {\it Rebound}, it is possible to input observed light curves then trace the outburst back in time to extract the initial conditions leading up to the outburst.

An open-source beta version of the {\it Rebound} ejecta dynamics package introduced in this paper will be released Spring 2021 on GitHub.

\section*{Acknowledgements}

The authors thank Dr. Hanno Rein at University of Toronto for his {\it Rebound} technical support in developing this package and Dr. Yanga Fernandez at University of Central Florida for his assistance in editing and reviewing the trigonometry of each system. We acknowledge support by the SSERVI Center for Lunar and Asteroid Science, which is run by UCF through NASA Grant 80NSSC19M0214.

\section*{Data Availability}
The data underlying this article will be shared on reasonable request to the corresponding author.




\bibliographystyle{mnras}
\bibliography{bibtex} 




\appendix
\section{Mathematical Derivations of Effects}
\label{appendixA}

Derivations of specific equations for each effect can be found here. Conceptual details and assumptions regarding values used in this study can be found in section \ref{sec:methods}. Similarly, specific implementations of effects using {\it Rebound} can be found in Appendix \ref{sec:appendixB}.

\subsection{Radiation Pressure Derivation}
Here we derive equation \ref{eq:aPR} that is used to describe the acceleration due to radiation pressure. We start with the total solar energy absorbed by a particle:
\begin{equation}
    \centering
    E_{total} = \epsilon K_{sc} \pi r^{2}
\end{equation}

\noindent
where $\epsilon$ is an absorption factor (here we assume $\epsilon = 1.5$, $K_{sc}$ is the solar constant calculated at the target body's distance from the sun, and $r$ is the particle radius. The solar constant is defined such that $K_{sc} = 1$ at 1 AU; however, since the amount of solar radiation decreases farther from the sun, we calculate $K_{sc}$ based on the target body's location relative to the sun. For systems close to 1 AU it is acceptable to approximate the solar constant as $K_{sc}$.

The power put out as a result of the solar radiation is described by

\begin{equation}
    \centering
    P_{rad} = m a c
\end{equation}

\noindent
where $m$ is the mass of the particle, $a$ is the acceleration that the particle experiences and $c$ is the speed of light. We set this equal to the total solar energy and solve for the acceleration, $a$.

\begin{equation}
    \centering
    a = \frac{\epsilon K_{sc} \pi r^{2}}{m c}
\end{equation}

\noindent
To simplify this expression, we can solve for the particle mass, $m = \frac{4}{3} \pi r^{3} \rho$, based on some $\rho$ that corresponds to the particle composition:

\begin{equation}
    \centering
    a = \frac{3 \epsilon K_{sc} \pi r^{2}}{4 \pi r^{3} \rho c}
\end{equation}

\noindent
Finally, simplifying this expression provides us with an acceleration expression in which the particle radius is the only varying quantity:

\begin{equation}
    \centering
    a = \frac{3 \epsilon K_{sc}}{4 r \rho c}
\end{equation}

\subsection{Ellipsoidal Gravitational Acceleration Derivation}
In this section, we derive the expression for the gravitational acceleration applied to particles around an ellipsoidal body. To do this we examine the derivation by \cite{Hu2017}, starting with the basic definition of an ellipsoid:

\begin{equation}
\centering
\frac{x'^{2}}{\lambda^{2}} + \frac{y'^{2}}{\lambda^{2} - h^{2}} + \frac{z'^2}{\lambda^2 - k^{2}} = 1    
\label{eq:ellipsoid}
\end{equation}

\noindent
where $\lambda$ represents the first solution of equation \ref{eq:ellipsoid} and $b = \sqrt{\lambda^{2} - h^{2}}$ and  $c = \sqrt{\lambda^{2} - k^{2}}$ are the semi-minor axes of the ellipsoid. Using this equation, we can calculate the semi-major axis of a reference ellipsoid proportional to the shape of the ellipsoidal body on which a particle is positioned. If the ellipsoidal body is denoted as ellipsoid 1 and the reference ellipsoid as ellipsoid 2, then $b_2 = \left( \frac{a_2}{a_1} \right) b_1$ and $c_2 = \left( \frac{a_2}{a_1} \right) c_1$. Given that $a_2 = \lambda$, $b_2 = \sqrt{\lambda^{2} - h^{2}}$, and $c_2 = \sqrt{\lambda^{2} - k^{2}}$, then 

\begin{equation}
\centering
\frac{1}{a_{2}^{2}} x'^{2} + \frac{ a_{1}^{2}}{a_{2}^{2} b_{1}^{2}} y'^{2} + \frac{a_{1}^{2}}{a_{2}^{2} c_{1}^2} z'^{2} = 1    
\label{eq:refellipsoid}
\end{equation}

\noindent
and

\begin{equation}
\centering
a_{2} = \sqrt{x'^{2} + \frac{ a_{1}^{2}}{b_{1}^{2}} y'^{2} + \frac{a_{1}^{2}}{c_{1}^2} z'^{2}}    
\label{eq:refsemimajor}
\end{equation}

\noindent
Equation \ref{eq:refsemimajor} describes the semi-major axis of the reference ellipsoid on which a point $\left( x', y', z' \right)$ rests in the local body coordinate frame. At this point we calculate the gravitational acceleration experienced at that point based on the gravitational potential. We define this potential using ellipsoidal harmonics to approximate the ellipsoidal nature of the body:

\begin{equation}
    \centering
    V = \sum_{n=0}^{\infty} \sum_{m=1}^{2n+1} c_{nm} \frac{F_{nm}(\lambda_{1})}{F_{nm}(a)} E_{nm}(\lambda_{2}) E_{nm}(\lambda_{3})
    \label{eq:potential}
\end{equation}

\noindent
where $c_{nm} = GM$ ($M$ is the mass of the large body), $E_{nm}$ is a Lam\'e function of the first kind, and $F_{nm}$ is a Lam\'e function of the second kind. For this approximation, we expand the Lam\'e functions of the first kind to the first order where $E{nm} \left( \lambda_{2} \right), E{nm} \left( \lambda_{3} \right) = 1$. Therefore, the following can also be derived from the definition of a Lam\'e function of the second kind given by \cite{Byerly1893}:

\begin{equation}
    \centering
    F_{nm} \left( x \right) = \left( 2m + 1 \right) \int_{\lambda}^{\infty} \frac{dx}{\sqrt{\left(x^{2} - h^{2} \right) \left(x^{2} - k^{2} \right)}}
    \label{eq:lame2}
\end{equation}

\noindent
Simplifying equation \ref{eq:potential} using equation \ref{eq:lame2} results in the following expansion \citep{Byerly1893}:

\begin{equation}
    \centering
    V =  M \int_{\lambda}^{\infty} \frac{dr}{\sqrt{\left(r^{2} - h^{2} \right) \left(r^{2} - k^{2} \right)}}
\end{equation}

\noindent
where M is the mass of the target body, and r represents the semi-major axis of the reference ellipsoid given in equation \ref{eq:refellipsoid}. Due to the dependency on the local particle position, the gradient is calculated in relation to the gradient of the semi-major axis.

\begin{equation}
    \centering
    \begin{cases}
    \frac{dV}{dx'} = \frac{M}{\sqrt{\left(a_{2}^{2}-h^{2}\right) \left(a_{2}^{2} - k^2\right)}} \frac{da_{2}}{dx'}
    \\
    \frac{dV}{dy'} = \frac{M}{\sqrt{\left(a_{2}^{2}-h^{2}\right) \left(a_{2}^{2} - k^2\right)}} \frac{da_{2}}{dy'}
    \\
    \frac{dV}{dz'} = \frac{M}{\sqrt{\left(a_{2}^{2}-h^{2}\right) \left(a_{2}^{2} - k^2\right)}} \frac{da_{2}}{dz'}
    \end{cases}
    \label{eq:gaccel}
\end{equation}

\noindent
where

\begin{equation}
    \centering
    \begin{cases}
    \frac{da_{2}}{dx'} = x' \left( x'^2 + \frac{a_{1}^{2}}{b_{1}^{2}} y'^{2} + \frac{a_{1}^{2}}{c_{1}^{2}} z'^{2} \right) ^{-\frac{1}{2}}
    \\
    \frac{da_{2}}{dy'} = \frac{a_{1}^{2}}{b_{1}^{2}} y' \left( x'^2 + \frac{a_{1}^{2}}{b_{1}^{2}} y'^{2} + \frac{a_{1}^{2}}{c_{1}^{2}} z'^{2} \right) ^{-\frac{1}{2}}
    \\
    \frac{da_{2}}{dz'} = \frac{a_{1}^{2}}{c_{1}^{2}} z' \left( x'^2 + \frac{a_{1}^{2}}{b_{1}^{2}} y'^{2} + \frac{a_{1}^{2}}{c_{1}^{2}} z'^{2}\right) ^{-\frac{1}{2}}
    \end{cases}
\end{equation}

\noindent
This acceleration is applied to each of the particles at their local locations. For testing purposes, we can examine this acceleration expression for a spherical body where $h, k = 0$ and $a_{1} = b_{1} = c_{1}$. Simplifying this expression assuming a spherical body does indeed produce the equation for the acceleration produced by a spherical body.

\subsection{Definition of Coordinate Systems}
In order to ensure the acceleration vectors are applied correctly in a system that moves in several directions (bodies orbiting, rotation along a tilted axis, particle interactions, etc.), we define two coordinate systems: a body-centric coordinate system (referred to hereafter as the local frame) and an ecliptic coordinate system (referred to hereafter as the global frame). While the general concepts of how these frames are related and how to switch between frames is referred to in section \ref{sec:methods}, here we describe the mathematical relations used to convert between the two frames.

\section{Computational Kinks and Performance Specific to {\it Rebound}}
\label{sec:appendixB}
\subsection{{\it Rebound} Specific Implementation}
While the Python version of {\it Rebound} is designed for straight forward implementation, developing such a small scale system involves several implementation tricks. An extensive description of specific implementations will be published with the Python package documentation.


For the implementation of ellipsoidal gravity, the default gravity built into the integrator must be eliminated. While it is possible to set the gravitational potential to "none" in {\it Rebound}, this is not recommended since the integration scheme requires a mass and gravitational potential in order to perturb particles properly. Since the particles experience a net acceleration due to all contributing forces in the system, we can counteract the default spherical gravitational acceleration on the particles by simply adding in an equal acceleration acting in the opposite direction to the default acceleration. After the spherical acceleration is eliminated, the ellipsoidal gravitational acceleration can be added using the add forces routine built into {\it Rebound}.

\subsection{Performance Evaluation}
\begin{figure}
    \centering
    \includegraphics[width=.47\textwidth]{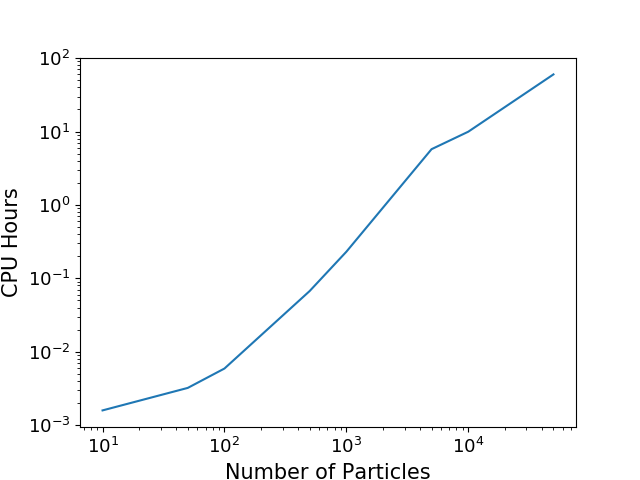}
    \caption{CPU run-time requirements for a basic Model A simulation with a varying number of particles and no additional effects.}
    \label{fig:cpu}
\end{figure}

Since we chose {\it Rebound} due to its improved computational performance, we benchmark the computational performance of the ejecta dynamics package by recording the CPU and clock run-time requirements for each simulation. Figure \ref{fig:cpu} depicts the increase in CPU requirements due to increasing number of particles. We determine that the maximum number of particles we are able to run in a single simulation is roughly $10^{5}$. Beyond this particle count, additional computational resources are required. A simulation of this magnitude requires approximately one week to run on one node composed of 24 cores. For simulations involving more than $10^{5}$ particles it is possible to run multiple simulations with $10^{5}$ particles and combine results. However, this method only works in simulations that do not involve particle-particle interactions. Due to the computational performance level of the ejecta dynamics package in {\it Rebound}, we are able to produce libraries of possible solutions for impacts given varying initial conditions. We are able to produce approximately 20 simulations per week given available resources on the STOKES computational cluster at UCF; therefore, a solution library of 200 solutions with varying initial conditions will take roughly 8-10 weeks to run, accounting for variations in cluster availability and maintenance. 

\bsp	
\label{lastpage}
\end{document}